\documentstyle[prabib,aps,psfig,12pt]{revtex}

\begin{document}

\title{Wormholes and Flux Tubes in the 7D Gravity on the Principal 
Bundle with SU(2) Gauge Group as the Extra Dimensions}
\author{V. Dzhunushaliev
\thanks{E-Mail Addresses : dzhun@rz.uni-potsdam.de and 
dzhun@freenet.bishkek.su; \qquad permanent address: 
Dept. Theor. Phys., Kyrgyz State National University, 
Bishkek 720024, Kyrgyzstan}}
\address{Institut f\"ur Mathematik, Universit\"at Potsdam 
PF 601553, D-14415 Potsdam, Germany} 
\author{H.-J. Schmidt
\thanks{http://www.physik.fu-berlin.de/\~{}hjschmi \ \ 
 \quad  hjschmi@rz.uni-potsdam.de}}
\address{Institut f\"ur Theoretische Physik, Freie
Universit\"at Berlin\\
and \\
Institut f\"ur Mathematik, Universit\"at Potsdam 
PF 601553, D-14415 Potsdam, Germany}

\maketitle

\begin{abstract}
Vacuum solutions for multidimensional gravity on the principal 
bundle with the SU(2) structural group as the extra 
dimensions are found and discussed. This  generalizes  
the results of Ref. \cite{vds3} from U(1) to the  SU(2) gauge group. 
The spherically symmetric solution 
with the off-diagonal components of the multidimensional 
metric   is obtained. It is shown that 
two types of solutions exist: the first has a wormhole-like 
4D base, the second is a gravitational flux tube with two 
color\footnote{Usually, this word refers to Quantum Chromodynamics
and its SU(3) group and not to SU(2). We, nevertheless, use that 
word here to bring the close analogy met between the
color degrees of freedom in both cases  to attention.}
  and electric charges. The solution depends  only on two 
parameters: the values of the electric and magnetic fields 
at the   origin. In the plane of these parameters there exists a curve 
separating the regions with different types of solutions. 
An analogy with the 5D solutions is discussed. 
\end{abstract}

\section{Introduction}

The presence of the off-diagonal components of the 
multidimensional (MD) metric can essentially change the 
properties of the appropriate MD gravitational equations. 
The physical reason for this is evident: these components 
of the MD metric are connected with the physical gauge fields 
(electromagnetic or Yang-Mills gauge fields) and excluding 
the physical fields 
can lead to this essential change.  
Let us repeat  the following 
theorem \footnote{which is useful for understanding 
how many degrees of freedom we have.} \cite{Sal1}, \cite{Per1}:
\par
Let $G$ be a structural group
of the principal  bundle.  Then  there  is a one-to-one
correspondence between the $G$-invariant metrics
\begin{equation}
ds^2 = \varphi (x^\alpha) \sum^{\dim G}_{a=5}
\left [\sigma ^a + 
A^a_\mu (x^\alpha)dx^\mu \right ]^2 +
g_{\mu\nu}(x^\alpha) dx ^\mu dx^\nu
\label{1-1}
\end{equation}
on the  total  space ${\cal X}$
and the triples $(g_{\mu \nu }, A^{a}_{\mu }, \varphi )$.
Here $g_{\mu \nu }$ is the 4D Einstein's pseudo  -
Riemannian metric on the base; $A^{a}_{\mu }$ are the gauge fields
of the group $G$ ( the nondiagonal components of
the multidimensional metric); $\varphi \gamma _{ab}$  is the
symmetric metric on the fibre 
($\sigma _{a} = \gamma _{ab}\sigma ^{b}, 
\gamma_{ab} = -\delta _{ab}$, 
$a=5, \dots ,$  dim $G$ is the index on the fibre and 
$\mu = 0,1,2,3$ is the index on the base).
\par 
According to this theorem we have the following independent 
degrees of freedom: the scalar field $\varphi (x^\alpha)$, 
the gauge fields $A^a_\mu (x^\alpha)$ and the 4D metric $g_{\mu\nu}(x^\alpha)$. 
Note that all fields in this MD gravity can depend only on 
the spacetime points (points on the base) as the total space 
is $G$-invariant. Such kind of MD gravity can easily  solve the 
following problem: why the physical degrees of freedom do not 
depend on the extra dimensions (ED). In addition to this, a 
 topological structure of the  ED is given that leads to 
a decrease of  the numbers of equations connected with the ED 
in comparison  with an ordinary MD gravity with non-fixed 
ED. Usually,  the number of these equations is too large: this  results in   
 essential problems with the compactification. Therefore,  
it is necessary to introduce some external fields in order 
to have a compactified ED. In our case the vacuum MD 
gravitational equations is sufficient to obtain the solutions 
with the compactified ED. In MD gravity on the principal 
bundle the preferable choice of the coordinate transformations 
\footnote{as the initial interpretation 
of Kaluza-Klein gravity} is  
\begin{eqnarray}
y'^a & = & y'^a (y^a) + f^a\left (x^\mu \right ) ,
\label{6}\\
x'^\mu & = & x'^\mu \left (x^\mu \right ) .
\label{7}
\end{eqnarray}
here $y^a$ are the coordinates on the fibre 
and $x^\mu$ are the coordinates on the base. 
The first term in (\ref{6}) means that the choice of coordinate 
system on the fibre is arbitrary. The second term indicates that 
in addition we can move the origin of the coordinate system on each 
fibre on the value $f^a(x^\mu)$.  
It is well known that such a 
transformation law   (\ref{6})  leads to a local gauge transformation for 
the appropriate non-Abelian field (see for overview 
\cite{overduin:1997pn}). 
That means,  the coordinate transformation  (\ref{6}) and (\ref{7}) 
are the most {\it natural} transformations for the MD 
gravitation on the principal bundle. Certainly we can perform  the 
much more general coordinate transformations: 
\begin{eqnarray}
y'^a & = & y'^a \left (y^a, x^\mu \right ) ,
\label{8}\\
x'^\mu & = & x'^\mu \left (y^a, x^\mu \right ) .
\label{9}
\end{eqnarray}
But in this case we will mix the points of the fibre 
\footnote{which are the elements of some group.} 
with the points of the base 
\footnote{which are the ordinary spacetimes points.}. 
Then the new coordinates $y'^a$ and $x'^\mu$ 
are not the coordinates along the fibre and the base of 
 the given bundle. 
\par
We can introduce the new coordinates and kill the 
$A^a_\mu$. But in this case the initial 4D metric can be 
changed very 
radical. For example, the initial static spherically symmetric 
4D metric can become nonstatic and nondiagonal. 
This situation is very clear in the initial Kaluza interpretation 
of 5D gravity with the constant and nonvariable 
$G_{55}$ component of the metric. In this case we have the 
ordinary vacuum electrogravity 
which is equivalent to the 5D gravity with the nondynamical 
$G_{55}$ component of MD metric. And we can choose the new 
coordinates $x'^B$ so that the initial 5D metric: 
\begin{equation}
ds^2 = \left (d\chi + A_\mu dx^\mu \right )^2 + 
g_{tt}dt^2 + g_{rr}dr^2 + r^2\left (d\theta ^2 + 
\sin ^2 \theta d\phi ^2 \right )
\label{1-2}
\end{equation}
will be 
\begin{equation}
ds^2 = G_{55}{d\chi '}^2 + g_{\mu\nu} dx'^\mu dx'^\nu
\label{1-3}
\end{equation}
Evidently metric (\ref{1-3}) will be nonstatic, nondiagonal 
and depend on the 5$^{th}$ coordinate. Of course,  the new coordinate 
system ${x'}^A$ ($A=0,1,2,3,5$) is worse than  the initial $x^A$. This 
remark allow us to say that the MD metric with the off-diagonal 
components of the metric $G^a_\mu$ 
can have  unusual properties in comparison  with the solutions 
where $G^a_\mu = 0$. 

\section{7D ansatz and equations}

Here we will consider the MD gravity on the principal bundle with SU(2) 
\footnote{this is the gauge group of the weak interaction} 
structural group. In this case the extra dimensions is 
SU(2) group \footnote{topologically it means that the ED are the 
$S^3$ sphere.} and the 4D physical spacetime is the base of this bundle. 
We will search a solution for the following 7D metric
\begin{equation}
ds^2 = \frac{\Sigma ^2(r)}{u^3 (r)} dt^2 - dr^2 - a(r) 
\left(d\theta^2 + \sin ^2\theta d\phi ^2 \right) - 
r_0^2 u(r)\left (\sigma ^a + A^a_\mu dx^\mu \right )^2 
\label{2-1}
\end{equation}
here $r_0$ is some constant, $\sigma ^a$ ($a=5,6,7$) are the 
Maurer-Cartan form with relation 
$d\sigma ^a = \epsilon^a_{bc} \sigma ^b\sigma ^c$
\begin{eqnarray}
\sigma ^{1} & = & {1\over 2}
(\sin \alpha d\beta - \sin \beta \cos \alpha d\gamma ),
\label{3-4-1}\\
\sigma ^{2} & = & -{1\over 2}(\cos \alpha d\beta +
\sin \beta \sin \alpha d\gamma ),
\label{3-4-2}\\
\sigma ^{3} & = & {1\over 2}(d\alpha +\cos \beta d\gamma ),
\label{3-4-3}
\end{eqnarray}
where $0\le \beta \le \pi , 0\le \gamma \le 2\pi , 
0\le \alpha \le 4\pi $ are the Euler angles. 
We choose the potential $A^a_\mu$ in the ordinary monopole-like form 
\begin{eqnarray}
A^{a}_{\theta } & = & {1\over 2}(1 - f(r))\{ \sin \phi ;-\cos \phi ;
0\} ,
\label{2-2}\\
A^{a}_{\phi } & = & {1\over 2}(1 - f(r))
\sin \theta \{\cos \phi \cos \theta ;
\sin \phi \cos \theta ;-\sin \theta \},
\label{2-3}\\
A^{a}_{t} & = & v(r)\{ \sin \theta \cos \phi ;
\sin \theta \sin \phi ;\cos \theta \},
\label{2-4}
\end{eqnarray}
Let us introduce the color electric $E^a_i$ and magnetic 
$H^a_i$ fields 
\begin{eqnarray}
E^a_i & = & F^a_{ti} ,
\label{2-5} \\
H^a_i & = & \sqrt \gamma \epsilon _{ijk} \sqrt {g_{tt}}F^{ajk}
\label{2-6}
\end{eqnarray}
here the field strength components are defined via 
$F^a_{\mu\nu} = A^a_{\nu ,\mu} - A^a_{\mu ,\nu} + 
\epsilon ^a_{bc}A^b_\mu A^c_\nu$, 
$\gamma$ is the determinant of the 3D space matrix, 
($i,j = 1,2,3$) are the space index. In our case we have
\begin{eqnarray}
E_r \propto v' , \qquad E_{\theta , \phi} \propto vf ,
\label{2-7} \\
H_r \propto \frac{\Sigma }{u^{3/2}}\frac{1 - f^2}{a} , 
\qquad H_{\theta , \phi} \propto f'
\label{2-8}
\end{eqnarray}
In order to have the wormhole-like (WH) solution we must demand that 
the functions $\Sigma (r), u(r), a(r), f(r)$ are  even functions and 
$v(r)$ is an  odd function. This means that at origin ($r=0$) 
we have only the radial $E_r$ and $H_r$ fields that indicate the 
presence of a flux tube of color electric and magnetic fields 
across the throat of this WH-like solution. The substitution to 
the 7D gravitational equations 
\footnote{The deduction of these gravitational equations for 
any  group $G$ is given in the Appendix.} 
\begin{eqnarray}
R^{\bar A}_\mu & = & 0
\label{2-9} \\
R^{\bar 5}_{\bar 5} + R^{\bar 6}_{\bar 6} + R^{\bar 7}_{\bar 7}
& = & 0 
\label{2-10} 
\end{eqnarray}
leads to the following system of equations 
\begin{eqnarray}
-R_{\bar 2\bar 2} - 2R_{\bar 3\bar 3} + R \propto  
\frac{\Sigma ''}{\Sigma} + \frac{a'\Sigma '}{a\Sigma} - 
\frac{4}{r_0^2u} - \frac{r_0^2u}{4a} f'^2 - 
\frac{r_0^2u}{8a^2}\left (f^2 -1 \right )^2 & = & 0 
\label{2-11} \\
R_{\bar 2\bar 2} - \frac{1}{2}R \propto 
24 \frac{\Sigma 'u'}{\Sigma u} - 24\frac{u'^2}{u^2} + 
16 \frac{a'\Sigma '}{a\Sigma} + 4 \frac{a'^2}{2a^2} - 
\frac{16}{a} + & & 
\nonumber \\
4 \frac{r_0^2u^4}{\Sigma^2}v'^2 - 
2\frac{r_0^2u}{a} f'^2 - 8 \frac{r_0^2u^4}{a\Sigma ^2}f^2v^2 + 
\frac{r_0^2u}{a^2}\left (f^2 -1 \right )^2 - 
\frac{48}{ur_0^2} & = & 0
\label{2-12}\\
R_{\bar 3\bar 3} = R_{\bar 4\bar 4} \propto  
\frac{a''}{a} + \frac{a'\Sigma '}{a\Sigma} - \frac{2}{a} + 
\frac{r_0^2u}{4a} f'^2 - \frac{r_0^2u^4}{a\Sigma ^2}f^2v^2 + 
\frac{r_0^2u}{4a^2}\left (f^2 -1 \right )^2 & = & 0
\label{2-13}\\
R^{\bar a}_{\bar a} \propto 
\frac{u''}{u} + \frac{u'\Sigma '}{u\Sigma} - 
\frac{u'^2}{u^2} + \frac{u'a'}{ua} - 
\frac{4}{r_0^2u} + & &  
\nonumber \\
\frac{r_0^2u^4}{3\Sigma^2}v'^2 - 
\frac{r_0^2u}{6a} f'^2 + 
\frac{2r_0^2u^4}{3a\Sigma ^2}f^2v^2 - 
\frac{r_0^2u}{12a^2}\left (f^2 -1 \right )^2 & = & 0
\label{2-15a} \\
R_{\bar 1\bar 5} \propto 
v'' + v'\left (-\frac{\Sigma '}{\Sigma} + 4\frac{u'}{u} + 
\frac{a'}{a}\right ) - \frac{2}{a} vf^2 & = & 0 ,
\label{2-14}\\
R_{\bar 3\bar 5} \propto 
f'' + f'\left (\frac{\Sigma '}{\Sigma} + 4\frac{u'}{u} \right ) 
+ 4\frac{u^3}{\Sigma ^2}fv^2 - \frac{f}{a} \left (f^2 -1 \right ) 
& = & 0 
\label{2-15} 
\end{eqnarray}
Note that the equations  (\ref{2-14}) and (\ref{2-15}) are the 
 ``Yang-Mills" equations. 
This system of ordinary differential  equations is extremely difficult 
for the analytical investigation 
\footnote{although there exists  a closed-form  solution for the simplest 
case when $a(r) = const$ (flux tube solution \cite{vds99}).}. 
Therefore we will search for a numerical solution of this system.

\section{Numerical Investigation}

Now we must write the initial conditions. At the origin ($r=0$) 
we can expand all functions by this manner 
\begin{eqnarray}
a(x) & = & 1 + \frac{a_2}{2}x^2 + \cdots ,
\label{2-16} \\
\Sigma (x) & = & \Sigma _0 + \frac{\Sigma _2}{2}x^2 + 
\cdots ,
\label{2-17}\\
u(x) & = & u_0 + \frac{u_2}{2}x^2 + \cdots ,
\label{2-18}\\
v(x) & = & v_1 x + \frac{v_3}{6}x^3 +\cdots ,
\label{2-19}\\
f(x) & = & f_0 + \frac{f_2}{2}x^2 + \cdots ,
\label{2-20}
\end{eqnarray}
here we introduce the dimensionless coordinate $x = r/\sqrt{a_0}$ 
($a_0 = a(0)$) and redefine $a(x)/a_0 \to a(x)$, 
$\sqrt{a_0} v(x) \to v(x)$, $r_0^2/a_0 \to r_0^2$. Then we can rescale 
time and the constant $r_0$ so that $\Sigma _0 = u_0 = 1$. Thus we 
have only the following initial conditions for the 
numerical calculations 
\begin{eqnarray}
a_0 = 1, \qquad u_0 = 1, \qquad \Sigma _0 = 1, \qquad v_0 = 0
, \qquad f_0 = f_0 ,
\label{2-21} \\
a'_0 = 0, \qquad u'_0 = 0, \qquad \Sigma '_0 = 0, \qquad v'_0 = v_1
, \qquad f'_0 = 0 ,
\label{2-22}
\end{eqnarray}
We see that our system depends on two parameters only: 
$f_0$ and $v_1$ \footnote{of course the 7D metric 
depends on the three parameters: $a_0$, $f_0$ and $v_1$.}. 
The constrained equation (\ref{2-12}) for the initial data 
give us 
\begin{equation}
r_0^2 = \frac{1 + \sqrt{1 + 3\left [v_1^2 + \frac{1}{4}
\left (f_0^2 - 1\right )^2\right ]}}{\left [v_1^2 + \frac{1}{4}
\left (f_0^2 - 1\right )^2\right ]}
\label{2-23}
\end{equation}
this equation is written  in  dimensionless variables. 
The numerical calculations of Eq.'s (\ref{2-11})-(\ref{2-15}) 
are presented in the Figs. 1-5. The initial data for these calculations 
are the following: 
$f_0=0.2, \quad v_1=0.3, \; 0.5, \; 0.6, \; 0.61, 
\; 0.615, \; 1.0, \; 2.0$. 
\par
\begin{figure}
\centerline{
\framebox{
\psfig{figure=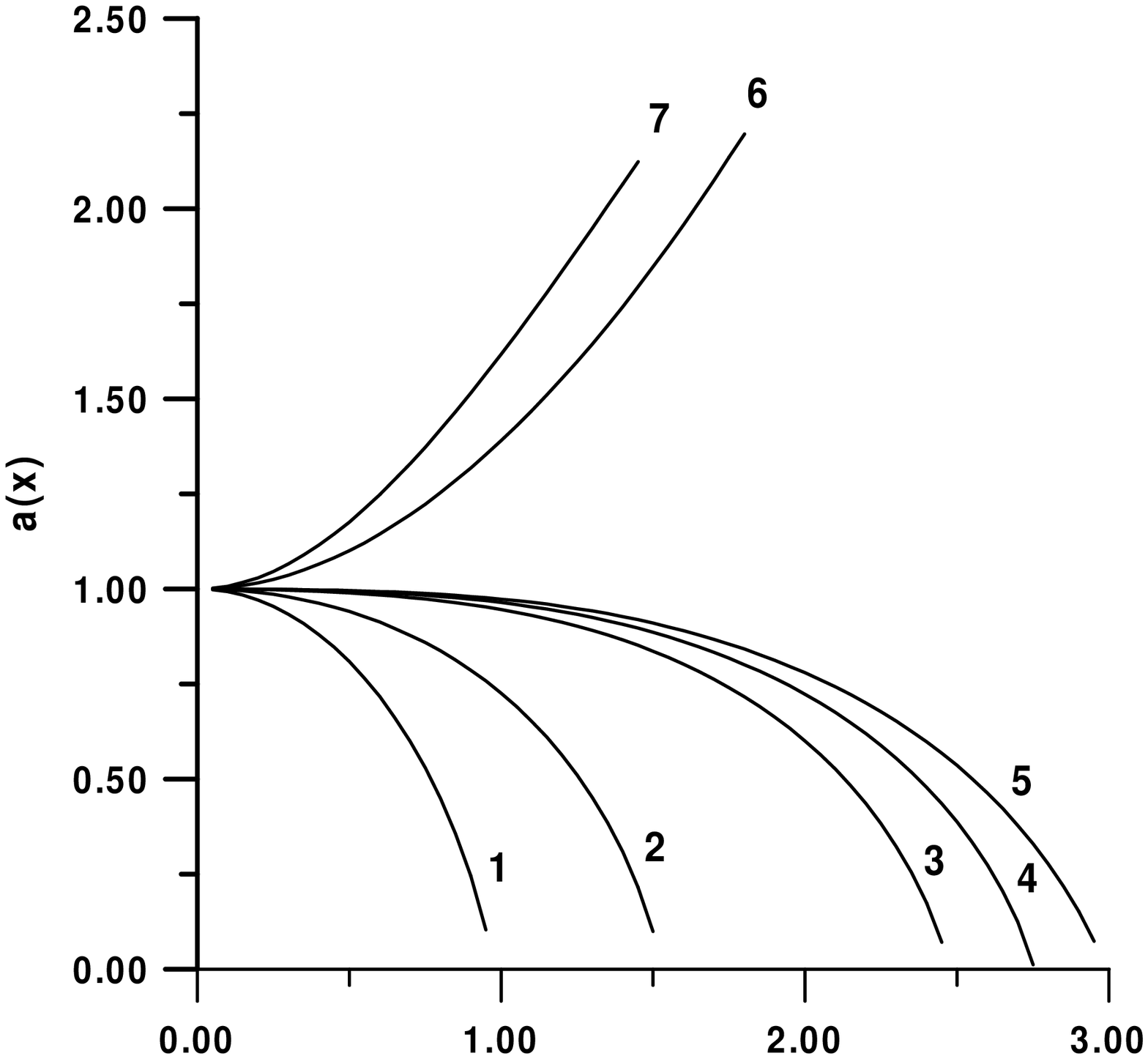,height=10cm,width=10cm}}}
\vspace{0.5cm}
\caption{Function $a(x)$. Initial dates: 
$f_0=0.2$, for curve 1:  $v_1=0.3$, 
for curve 2 : $v_1=0.5$, 
for curve 3:  $v_1=0.6$, 
for curve 4:  $v_1=0.61$, 
for curve 5:  $v_1=0.615$, 
for curve 6:  $v_1=1.0$, and
for curve 7:  $v_1=2.0$.} 
\label{fig1}
\end{figure}
\begin{figure}
\centerline{
\framebox{
\psfig{figure=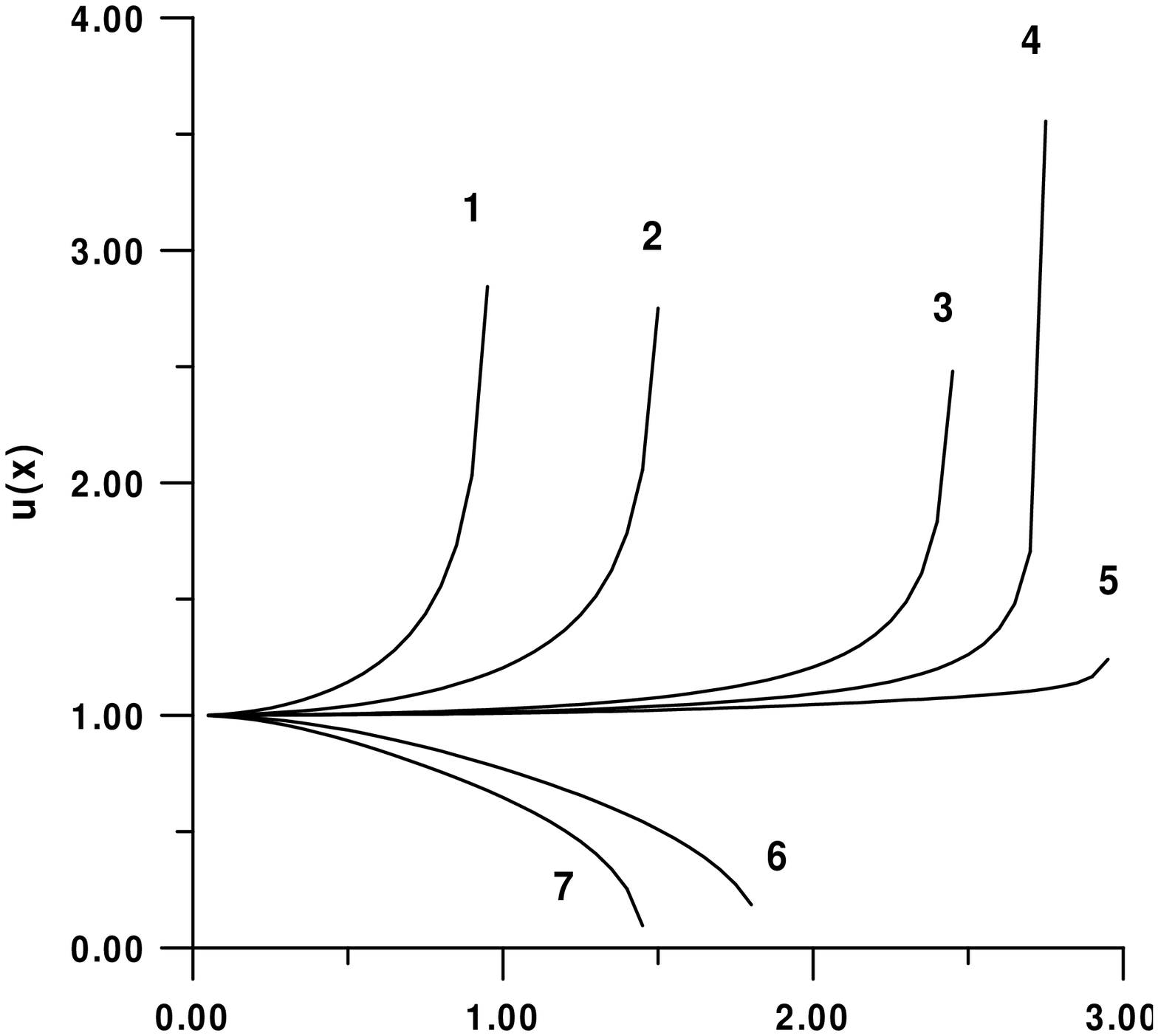,height=10cm,width=10cm}}}
\vspace{0.5cm}
\caption{Function $u(x)$. Initial dates as for Fig. 1.} 
\label{fig2}
\end{figure}
\begin{figure}
\centerline{
\framebox{
\psfig{figure=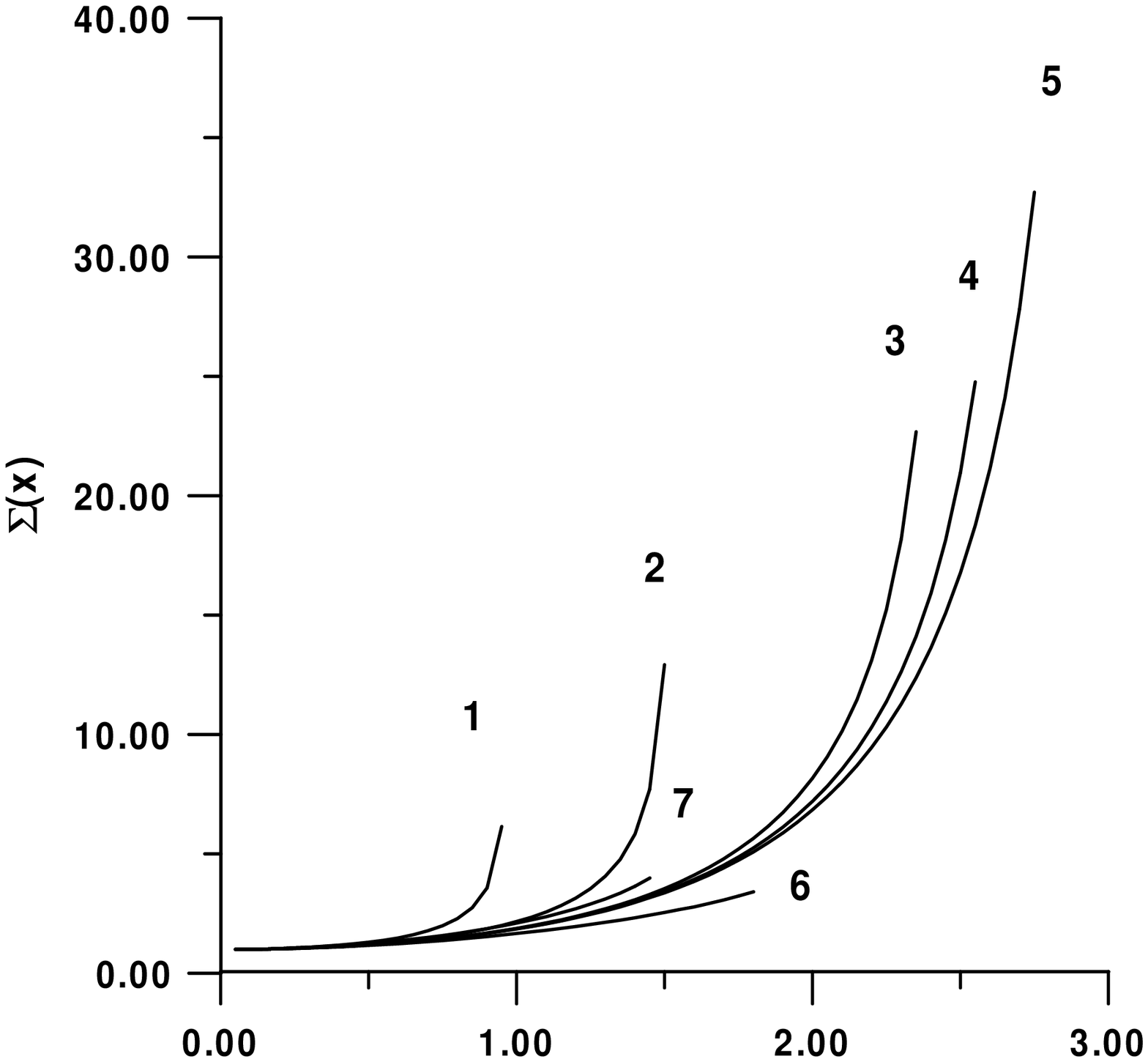,height=10cm,width=10cm}}}
\vspace{0.5cm}
\caption{Function $\Sigma(x)$. Initial dates as for Fig. 1.} 
\label{fig3}
\end{figure}
\begin{figure}
\centerline{
\framebox{
\psfig{figure=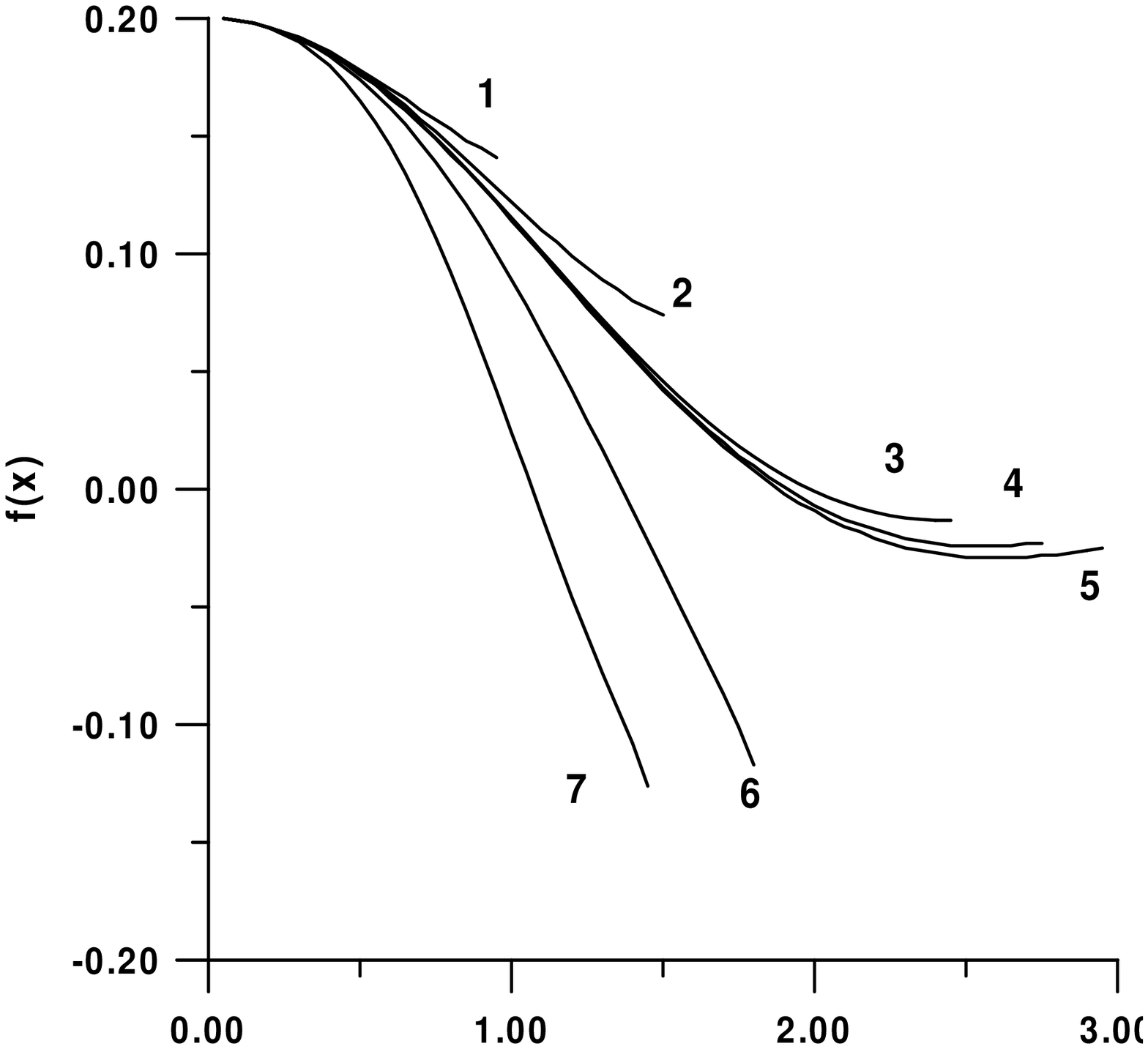,height=10cm,width=10cm}}}
\vspace{0.5cm}
\caption{Function $f(x)$. Initial dates as for Fig. 1.} 
\label{fig4}
\end{figure}
\begin{figure}
\centerline{
\framebox{
\psfig{figure=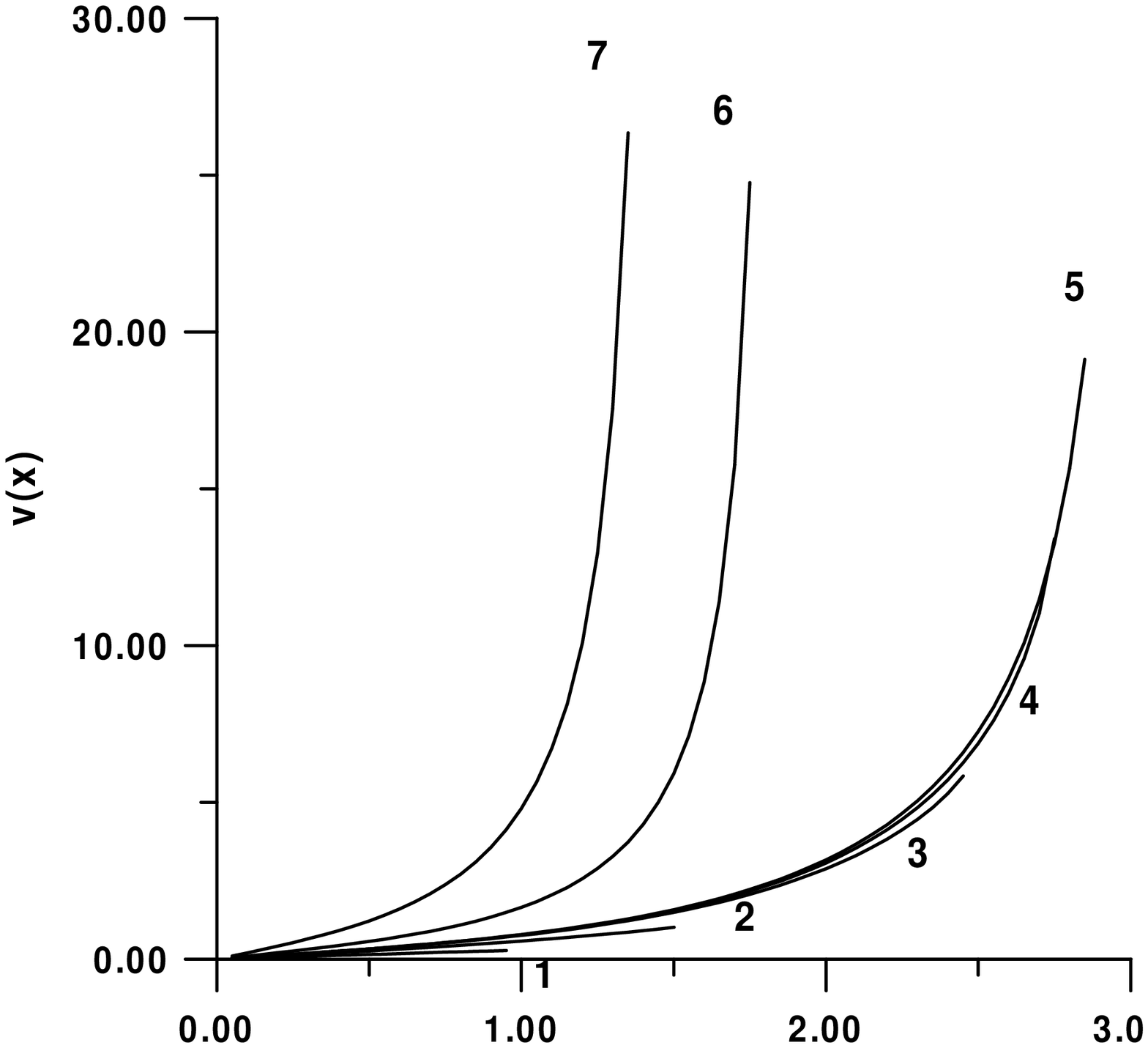,height=10cm,width=10cm}}}
\vspace{0.5cm}
\caption{Function $v(x)$. Initial dates as for Fig. 1.} 
\label{fig5}
\end{figure}

\section{Physical discussion}

Figs. 1-5 show us that in accordance with some relation between 
$f_0$ and $v_1$ there are  two types of solutions: 
\begin{enumerate}
\item
There exists some value of the radial coordinate $r_1$ that 
$a(\pm r_1) = 0$, $u(\pm r_1) = s(\pm r_1) = \infty$. 
Probably this means that at $r = \pm r_1$ points 
we have a singularity. This 4D part of the MD metric 
we can name as gravitational flux tube. As we have 
some flux of color electric/magnetic field between the points 
$r = -r_1$ and $r = +r_1$ where the color electric/magnetic 
charges are located.  
\item
There exists some value of the radial coordinate $r_2$ that 
$a(\pm r_2), s(\pm r_2) < \infty$, $u(\pm r_2) = 0$. 
The value $u(\pm r_2) = 0$ means that the interval 
$ds^2 = 0$ on the hypersurface 
$r = \pm r_2, t,\theta ,\phi = const$. Since the value 
of $a(\pm r_2)$ is finite we can name this type of the solutions 
as the WH-like. 
\end{enumerate}
\par 
This  type of the solutions is close to 
the 5D case investigated in \cite{vds3}. 
In this Ref. it was shown that the 5D vacuum Kaluza-Klein 
theory for the metric 
($r_0$ is radius of the U(1) gauge group, $Q = nr_0$ 
is the magnetic charge, $\chi$ is the 5$^{th}$ coordinate)
\begin{eqnarray}
ds^2 &=& e^{2\nu (r)}dt^{2} - r_0^2e^{2\psi (r) - 2\nu (r)}
\left [d\chi +  \omega (r)dt + n\cos \theta d\phi \right ]^2
\nonumber \\
&-& dr^{2} - a(r)(d\theta ^{2} +
\sin ^{2}\theta  d\phi ^2),
\label{3-1}
\end{eqnarray}
has the following solutions 
\begin{enumerate}
\item
{\bf E $>$ H} \   \footnote{$E = q/a$ and $H = Q/a$ are the electric and 
magnetic charges.}. WH-like solution located between two $ds^2 = 0$ 
hypersurfaces.
\item
{\bf E = H}. Infinite flux tube with constant electric and 
magnetic fields. 
\item
{\bf E $<$ H}. Finite flux tube located between two singularities 
at points  $r = \pm r_1$  where are the electric and 
magnetic charges.
\end{enumerate}
In the first 5D case the solution exists for $|r| > r_0$ 
where $e^{2\nu} < 0$ and the metric is asymptotically flat. 
The whole construction can be interpreted as the Euclidean 
WH with the Lorentzian throat  
\footnote{The words Euclidean and Lorentzian can be exchanged.} 
\cite{vdschmidt2}. The question is: whether  this 
situation can be kept in the  7D case for the second type of the solutions? 
In Ref's \cite{bronnikov95a}, \cite{bronnikov95b} a similar   
idea had been investigated about the changing of the signature 
of the 4D metric in some MD metric on the regular $T$-hole 
horizon. 
\par 
Also in our case the numerical calculations show that on the 
$(f_0, v_1)$ plane there is a curve that separates regions with 
the different solution type. Evidently, in the first rough 
approximation the equation for this curve is
\begin{equation}
\frac{a''_0}{a_0} = \frac{2}{a_0} - \frac{r_0^2 u_0}{a_0^2} 
\left (f_0^2 - 1 \right )^2 = 0 
\label{3-2}
\end{equation}
In  Fig. 6 these regions with the  different type 
of solutions are shown. 
\begin{figure}
\centerline{
\framebox{
\psfig{figure=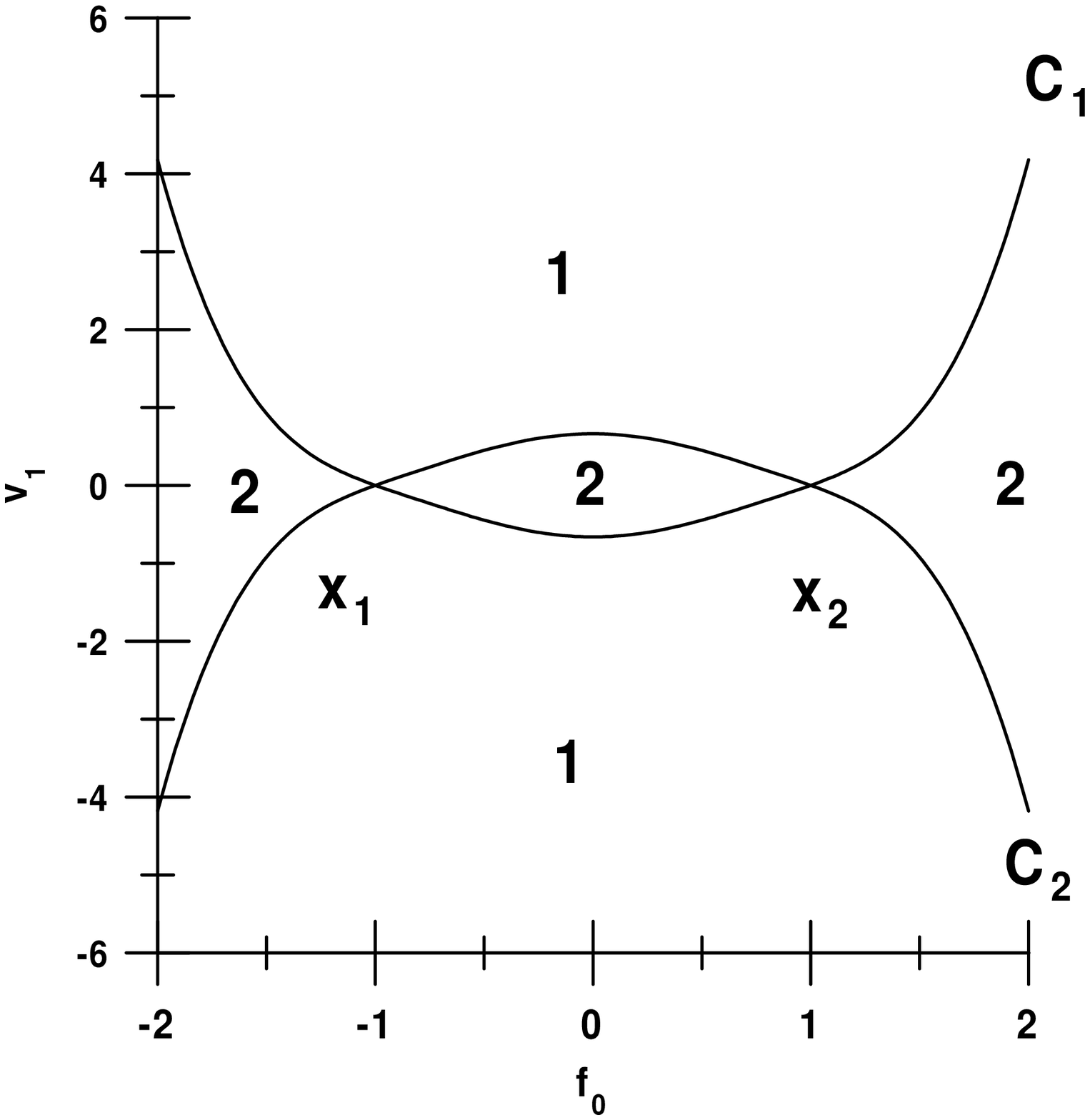,height=10cm,width=10cm}}}
\vspace{0.5cm}
\caption{The curves C$_1$ and C$_2$ separate  the regions with different 
types of the solutions. In the 2-regions with $a_0'' < 0$ 
we have the flux tube solutions and for the 1-regions $a_0'' > 0$ - 
the wormhole-like solutions.} 
\label{fig6}
\end{figure}
\par 
In the $(f_0, v_1)$ plane we can single out a few cases 
allowing a  more detailed analysis. 

\subsection{$f_0 = \pm 1, v_1 = 0$}

Immediately  from   the  ``Yang-Mills'' equations (\ref{2-14}), 
(\ref{2-15}) we see that 
\begin{equation}
v(r) = 1 , \qquad f(r) = \pm 1
\label{3-3}
\end{equation}
All terms with gauge fields in (\ref{2-11})-(\ref{2-15a}) 
vanish,  and at the  origin $(r = 0)$ the equation for the  initial data  
give us 
\begin{equation}
\frac{1}{a_0} + \frac{3}{u_0r_0} = 0 
\label{3-4}
\end{equation}
As $a(r), u(r) > 0$ this relation cannot be satisfied. 
This allows us to say that {\it in the  absence of the gauge 
fields (off-diagonal components of the MD metric)
the gravitational flux tubes and WH-like solutions 
do not exist}. 

\subsection{$f = 0$}

In this case (\ref{2-14}) equation is easy to integrate 
\begin{equation}
v' = \frac{q \Sigma}{r_0^2 a u^4} 
\label{3-5}
\end{equation}
And from (\ref{2-7}), (\ref{2-8}) we see that there are 
only the radial color component of the electric and 
magnetic fields. 
\par It is very interesting to note that there is some 
very simple analytical solution in the case $a(r) = const$ 
\cite{vds99}. The solution is 
\begin{eqnarray} 
a(r) &=& {2 q^2 \over 7} = {r_0 ^2 \over 8} = const 
\label{3-6} \\
\Sigma (r) &=& \cosh \left( {7 r \over 2 \sqrt{2} q} \right)  
\label{3-7} \\
v(r) &=& {\sqrt{2} \over r_0} \sinh \left( {7 r \over 2 \sqrt{2} q} \right) 
\label{3-8} \\
u(r) & = & 1 
\label{3-9} \\
q & = & \sqrt\frac{7a}{2} 
\label{3-10} 
\end{eqnarray} 
We can apply the following gauge transformation to the 
potentials of Eqs. (\ref{2-2} - \ref{2-4}) with $f(r) = 0$
\begin{equation} 
\label{3-11} 
A'_{\mu} = S^{-1} A_{\mu} S - i (\partial _{\mu} S^{-1} )S 
\end{equation} 
where 
\begin{equation} 
\label{3-12} 
S =  \left ( 
\begin{array} {cc} \cos {\theta \over 2} & 
-e^{-i \phi} \sin {\theta \over 2} \\ 
e^{i \phi} \sin {\theta \over 2} & 
\cos {\theta \over 2} 
\end{array} \right ) 
\end{equation} 
With this gauge transformation we find that the gauge 
potentials become 
\begin{eqnarray} 
\label{3-13} 
A'^a _{\theta} &=& (0; 0 ; 0) \\ 
\label{3-14} 
A'^a _{\varphi} &=& (\cos \theta -1 ) (0 ; 0; 1) \\ 
\label{3-15} 
A'^a _t &=& v(r) (0; 0 ; 1) 
\end{eqnarray} 
In fact this means that the potential  (\ref{3-13})-(\ref{3-15}) 
is an Abelian one. 
This solution  corresponds  to the points $x_1$ and $x_2$ 
in  Fig. 6 and we can call it  an 
{\it infinite gravitational flux tube.} It is very close to 
the Levi-Civita-Robinson-Bertotti solution 
\cite{levi-civita17}, \cite{robinson59}, \cite{bertotti59} 
in 4D electrogravity. This remark poses  a very interesting 
problem: how are  the other solutions on the curve $C_1$ $C_2$ 
(Fig. 6) ? Are these solutions infinite flux tubes  or something else
\footnote{Unfortunately, here the numerical investigation 
is more difficult because of a sensitive dependence on initial data.
 This fact becomes also clear by the following consideration: 
These lines are the borderlines between two different domains of
 attraction, and there this sensitive behaviour is quite typical.} ? 
\par 
For the question which other Lie groups can lead to 
similar results we can say that for any $N > 2$ there  is an inclusion 
SU$(2) \subset {\rm SU}(N)$. This means that for any such  SU$(N)$  we can 
use the ansatz   (\ref{2-2})-(\ref{2-4}).  As one knows, also all 
other compact semi-simple Lie groups (besides U(1), of course),
 have SU(2) as one of their subgroups. Therefore, for all these Lie groups 
similar statements are also valid.  For the other Lie 
groups it is  complicated to  find out the spherically 
symmetric ansatz for the off-diagonal components of the 
MD metric. 
\par 
We would like to point out the following dimensional reduction 
expression for the Ricci scalar $R(E)$ on the total space 
of the principal bundle 
\begin{eqnarray}
\int d^4 x d^d y \sqrt{ \vert \det G_{AB} \vert} R(E) = 
\nonumber \\
V_G \int d^4 x \sqrt{ \vert  g \vert } \varphi ^{d/2} 
\left [ R(M) + R(G) - \frac{1}{4} \varphi F^a_{\mu\nu}F_a^{\mu\nu} 
+ \frac{1}{4}d (d - 1) \partial _\mu \varphi \partial ^\mu \varphi 
\right ]
\label{3-16}
\end{eqnarray}
here $R(M), R(G)$ are the Ricci scalars of the base and structural group 
of the principal bundle respectively; $V_G$ is the volume of the group
$G$, cf. \cite{Per1}, eq. (8.12). 
Immediately we see that 
for an arbitrary Lie group $G$ we face the  problem of how to get the
appropriate  ansatz for  the gauge field $A^a_\mu$.

In this context it should be noted that also in the models discussed here, a 
generalized Birkhoff theorem (cf. \cite{schx}) is valid. This means that
a spherically symmetric solution possesses a further isometry without
 additional assumptions. Of course, this  result rests  on the symmetries 
 chosen for our model.

\section{The possible physical applications}

Probably, the most interesting above-mentioned solution 
in 7D gravity on the principal bundle with the SU(2) 
structural group is the WH-like solution. We remember that 
if the $G_{55},G_{66},G_{77}$ components of the MD metric 
are not the dynamical variables then this MD gravity is 
equivalent to the pure 4D gravity + SU(2) Yang-Mills 
theory. It can be supposed that in the Universe there exist regions 
where the $G_{55},G_{66},G_{77}$ metric components are  
nondynamical variables and that there  exist other  regions where  
these components are dynamical variables. The composite 
WH can be of such a  kind.  It means that we have 
our above-mentioned WH-like solution as a throat
of composite WH and two 4D Yang-Mills black holes attached to this 
throat  (see Fig. 7). 
\begin{figure}
\centerline{
\framebox{
\psfig{figure=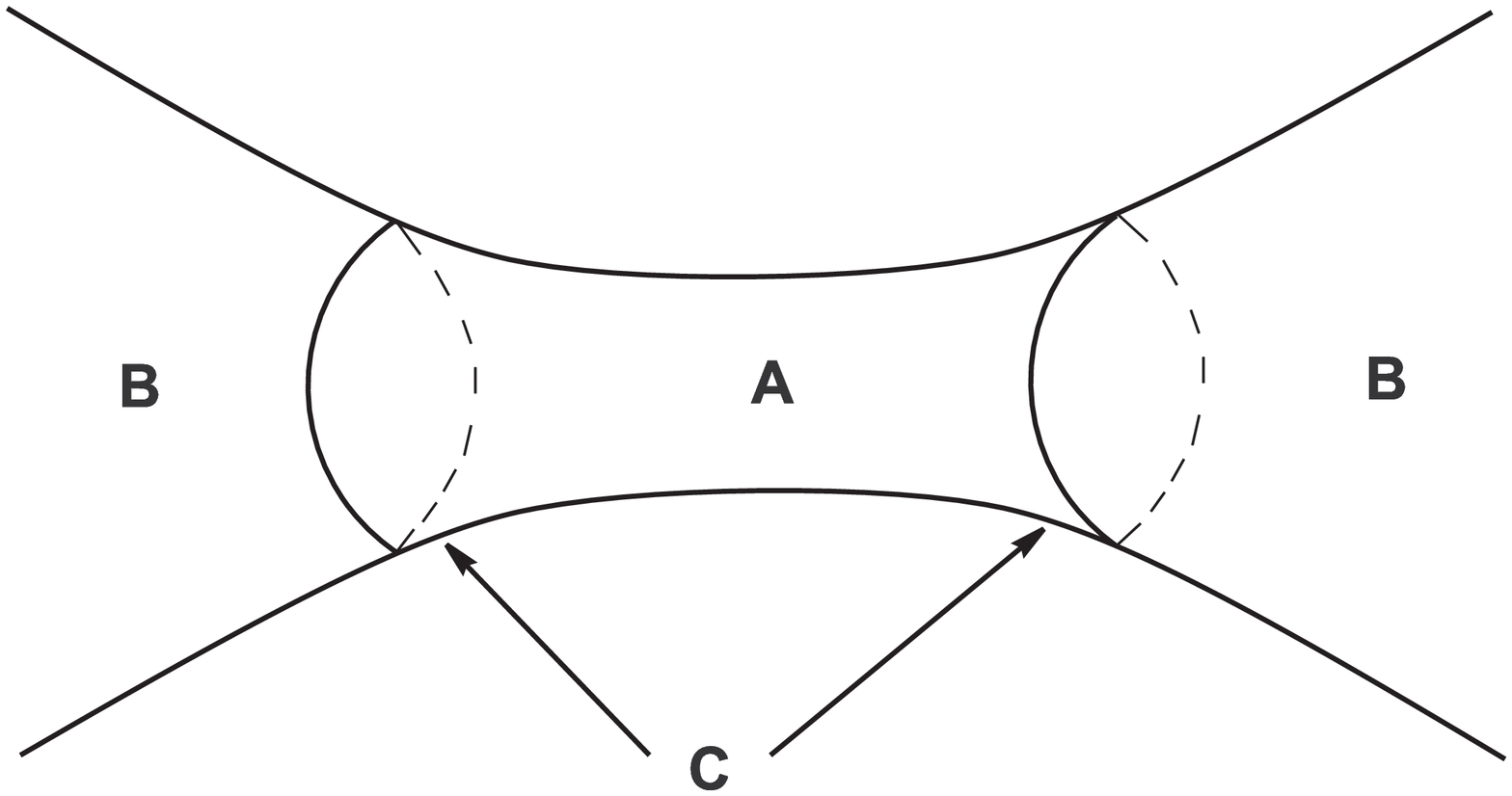,height=6cm,width=10cm,angle=-90}}}
\vspace{0.5cm}
\caption{The composite wormhole with the 1-throat as MD wormhole-like 
solution and two black 2-holes attached to its 3-ends.} 
\label{fig7}
\end{figure}
\par 
Such construction can polarize a space-time foam by the 
following way \cite{dzh7}. Without presence of the 7D (5D) 
throat the handles of the space-time foam are located 
disordered  in the space-time (Fig. 8a). But after the 
appearance of the throat of composite WH the location 
of these handles is ordered (Fig. 8b). Such model 
can be a geometrical model of the renormalization of the 
color/electric charge (the case for the electric  charge 
is described in \cite{dzh7}). This composite WH with 
the polarization of the space-time foam is a continuation 
of Wheeler's  idea about ``charge without charge'' and 
``mass without mass'' in the {\it vacuum} gravity. 
\begin{figure}
\centerline{
\framebox{
\psfig{figure=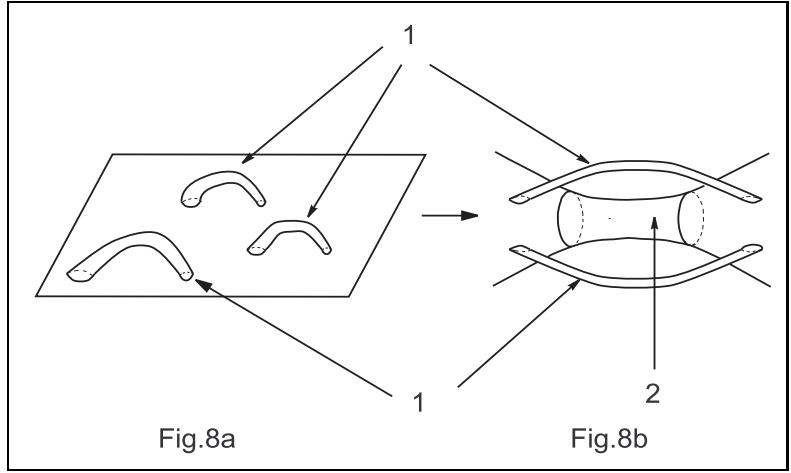,height=6cm,width=10cm,angle=-90}}}
\vspace{0.5cm}
\caption{A polarization of space-time foam in the presence of MD
insertion. Fig. 8a presents the unpolarized space-time foam. 
Fig. 8b presents the polarized space-time foam. 
1 are the virtual wormholes, 
2 is the multidimensional insertion.} 
\label{fig8}
\end{figure}
\par 
In this connection we can recall Wheeler's quote 
in Ref. \cite{wheel1}: ``We are therefore led to consider 
the view that the electron is nothing but a collective state of 
excitation of the foam-like medium $\ldots$ In other 
words the electron is not a natural starting point for the 
description of nature, according to the present 
reinterpretation of the views of Lorentz. Instead 
it is a first order correction to vacuum physics. That 
vacuum, that zero order state of affairs, with its enormous 
concentrations of electromagnetic energy and multiply-connected 
topologies, has to be described properly before one has the 
starting point for a proper perturbation theoretic development.'' 
May be all these words can be used to a neutrino with mass, 
i.e. neutrino is a wormhole in the polarized space-time foam. 
In these cases there remains to solve the  problem of  the geometric description 
of spin.  A similar  6-dimensional model, called $M$-fluxbranes, 
has been discussed recently in \cite{galtx}. 
\par 
It is interesting to note that the above-mentioned 
gravitational flux tube solutions cannot be the 
throat of the composite WH as they have a singularity on the 
place of the $ds^2=0$ hypersurface for the  WH-like solution. 
It imposes some restriction on the possible relation between 
color electric  and magnetic fields in the composite WH.

\section*{Acknowledgements}
We would like to thank A. Kirillov and B. Schmidt for useful comments. 
VD is grateful for financial support by a Georg Forster 
Research Fellowship from the Alexander von Humboldt Foundation. 
 HJS thanks the DFG for financial support.

\appendix

\section{Gravitational equations for the gravity on the principal 
bundle}

We  consider the MD gravity on the principal bundle with 
the  structural group $G$ \cite{dzh2}. 
In this case the extra dimensions are the  
 group $G$,  and the 4D physical spacetime is the base of this bundle. 
\par 
According to above-mentioned theorem the MD metric on the 
total space is
\begin{equation}
ds^2 = \Sigma _{\bar A} \Sigma ^{\bar A} 
\label{a-1}
\end{equation}
where 
\begin{eqnarray}
\Sigma ^{\bar A} & = & h^{\bar A}_B dx^B ,
\label{a-2} \\
\Sigma ^{\bar a} & = & \varphi(x^\alpha)\sigma ^{\bar a} + 
h^{\bar a}_\mu (x^\alpha)dx^\mu ,
\label{a-3} \\
\sigma ^{\bar a} & = & h^{\bar a}_b dx^b , 
\label{a-3a} \\
\Sigma ^{\bar \mu} & = & h^{\bar \mu}_\nu (x^\alpha)dx^\nu 
\label{a-4}
\end{eqnarray}
here $x^B$ are the coordinates on the total space 
($B = 0,1,2,3, 5, \dots ,N$, is the MD index dim $(G) = N$), 
$x^a$ is the coordinates on the group $G$ 
($a = 5, \dots , N$), 
$x^\mu = 0,1,2,3$ are the coordinates on the base of the bundle, 
$\bar A$ is the $N$-bein index, $h^{\bar A}_B$ is the $N$-bein, 
$\sigma ^{\bar a} $ are the 1-forms on the  group $G$ satisfying  
$d \sigma ^{\bar a} = f^a_{bc}\sigma ^{\bar b} \sigma ^{\bar c} $, 
($f^a_{bc}$) are the structural constants for the  group $G$. 
We must note that the functions $\varphi,h^{\bar a}_\mu, h^{\bar \mu}_\nu$
can depend only on the $x^\mu$ points on the base 
as  the fibres of  our bundle  
are locally homogeneous spaces.
The matrix $h^{\bar A}_B$ has the following form
\begin{equation}
h^{\bar A}_B = 
\left (
\begin{array}{cc}
\varphi h^{\bar a}_b & h^{\bar a}_\mu \\
0 & h^{\bar \nu}_\mu
\end{array}
\right )
\label{a-5}
\end{equation}
The inverse matrix $h_{\bar A}^B$ is 
\begin{equation}
h_{\bar A}^B = 
\left (
\begin{array}{cc}
\varphi ^{-1}h_{\bar a}^b & h_{\bar \mu}^b \\
0 & h_{\bar \mu}^\nu
\end{array}
\right )
\label{a-6}
\end{equation}
here $h_{\bar \mu}^b = -\varphi ^{-1} h^b_{\bar a}
h^{\bar a}_\nu h^\nu_{\bar \mu}$. 
Also we see that we have only the following degrees of freedom: 
$\varphi (x^\alpha), h_{\bar \mu}^b(x^\alpha)$ and 
$h_{\bar \mu}^\nu(x^\alpha)$, $h_{\bar b}^a$ is given and not 
varying. Varying with respect to 
$h^A_{\bar\mu} = (h^a_{\bar\mu}, h_{\bar \mu}^\nu)$ leads to the equations 
\begin{equation}
R^{\bar \mu}_A - \frac{1}{2} h^{\bar \mu}_A R = 0
\label{a-7}
\end{equation}
here $\bar A = \bar a$, $\bar \nu$, $R^{\bar A}_B$ is 
the MD Ricci tensor. Let $x^a$ be  the coordinates on the 
 group $G$ then 
\begin{equation}
\varphi \sigma ^{\bar a} = \varphi h^{\bar a}_b dx^b 
\label{a-8}
\end{equation}
Varying with respect to $\varphi (x^\mu)$ leads to the following 
result
\begin{equation}
\frac{\delta }{\delta \varphi} 
\left (h R \right ) = 
\frac{\delta \left (h_{\bar a}^b/\varphi \right )}{\delta \varphi} 
\frac{\delta \left (h R \right )}
{\delta \left (h_{\bar a}^b/\varphi \right )} = 
- \frac{1}{\varphi ^2} h_{\bar a}^b 
\left (R^{\bar a}_b - \frac {1}{2} h^{\bar a}_b R  \right ) = 0 
\label{a-9}
\end{equation}
here $h = \det h_B^{\bar A}$, $R$ is the MD Ricci scalar for the 
metric on the total space. As $h_{\bar a}^\nu = 0$ we can write 
\begin{equation}
h^b_{\bar a}\left (R^{\bar a}_b - \frac{1}{2}h^{\bar a}_b R \right ) + 
h^\nu_{\bar a}\left (R^{\bar a}_\nu - \frac{1}{2}h^{\bar a}_\nu R \right ) = 
h_{\bar a}^A 
\left (R^{\bar a}_A - \frac {1}{2} h^{\bar a}_A R  \right ) = 0 
\label{a-9a}
\end{equation}
From (\ref{a-7}) and (\ref{a-9a}) we see that 
\begin{equation}
h_{\bar a}^A \left (R^{\bar a}_A - 
\frac {1}{2} h^{\bar a}_A R  \right ) +  
h_{\bar \mu}^A
\left (R^{\bar \mu}_A - \frac {1}{2} h^{\bar \mu}_A R  \right ) = 
h^A_{\bar B}\left (R^{\bar B}_A - 
\frac{1}{2} h^{\bar B}_A R \right ) = 0 
\label{a-10} .
\end{equation}
This means that
\begin{equation}
R = 0
\label{a-11} .
\end{equation}
Hence from (\ref{a-9a}) we can write 
\begin{equation}
h^A_{\bar a}R^{\bar a}_A = R^{\bar a}_{\bar a}
\label{a-14}
\end{equation}
Finally we have the following equation system for the MD 
gravity on the principal bundle 
\begin{eqnarray}
R^{\bar \mu}_A & = & 0
\label{a-15} \\
R^{\bar a}_{\bar a} = R^{\bar 5}_{\bar 5} + 
\cdots + R^{\bar N}_{\bar N} & = & 0 .
\label{a-16} 
\end{eqnarray}
We note that the (\ref{a-16}) Eq. is an analog for the 
Brans-Dicke scalar gravity. 
\par 
In addition we see that (\ref{a-15}) can be wrote as 
\begin{equation}
R_{\bar\mu A} = 0 \quad or \quad 
h^B_{\bar A} R_{\bar\mu B} = R_{\bar\mu \bar A} = 0
\label{a-17}
\end{equation}

%\bibliography{john,john2}
%\bibliographystyle{prsty}

\newpage

{\bf 
Figure captions}
\par
Fig. 1. Function $a(x)$. Initial data: 
$f_0=0.2$, for curve 1:  $v_1=0.3$, 
for curve 2 : $v_1=0.5$, 
for curve 3:  $v_1=0.6$, 
for curve 4:  $v_1=0.61$, 
for curve 5:  $v_1=0.615$, 
for curve 6:  $v_1=1.0$, and
for curve 7:  $v_1=2.0$. 
\par
Fig. 2. Function $u(x)$. Initial data as for Fig. 1.
\par
Fig. 3. Function $\Sigma(x)$. Initial data as for Fig. 1.
\par
Fig. 4. Function $f(x)$. Initial data as for Fig. 1.
\par
Fig. 5. Function $v(x)$. Initial data as for Fig. 1.
\par
Fig. 6. The curves C$_1$ and C$_2$ separate  the regions with different 
types of the solutions. In the 2-regions with $a_0'' < 0$ 
we have the flux tube solutions and for the 1-regions $a_0'' > 0$ - 
the wormhole-like solutions.
\par
Fig. 7. The composite wormhole with the 1-throat as MD wormhole-like 
solution and two black 2-holes attached to its 3-ends.                            
\par
Fig. 8. A polarization of space-time foam in the presence of MD
insertion. Fig. 8a presents the unpolarized space-time foam. 
Fig. 8b presents the polarized space-time foam.
1 are the virtual wormholes, 
2 is the multidimensional insertion.

\end{document}